\documentclass[prd,aps,onecolumn,superscriptaddress,nofootinbib,showpacs,a4paper,notitlepage,10pt]{revtex4-1}

\usepackage{amsmath,amscd,amssymb,amsfonts}
\newcommand\id{\leavevmode\hbox{\small1\kern-3.3pt\normalsize1}}
\usepackage{graphicx}
\usepackage{epstopdf}
\usepackage{geometry}
\usepackage{psfrag}
\usepackage{epsfig}
\usepackage{cancel}
\usepackage{textcase}
\usepackage{xcolor}

\newcommand\ii{\leavevmode\hbox{\small1\kern-3.3pt\normalsize1}}

\newcommand{\bra}{\langle}
\newcommand{\ket}{\rangle}

\newcommand{\qed}{\nobreak \ifvmode \relax \else
          \ifdim\lastskip<1.5em \hskip-\lastskip
          \hskip1.5em plus0em minus0.5em \fi \nobreak
          \vrule height0.75em width0.5em depth0.25em\fi}

\begin {document}
\title{Entanglement between smeared field operators in the Klein-Gordon vacuum}
\author{Magdalena Zych}
\affiliation{Department of Theoretical Physics II, University of \L{}\'{o}d\'{z}, ul. Pomorska 149/153, \L{}\'{o}d\'{z}, Poland}

\author{Fabio Costa}
\author{Johannes Kofler}
\author{\v{C}aslav Brukner}
\affiliation{Faculty of Physics, University of Vienna, Boltzmanngasse 5, Vienna, Austria}
\affiliation{Institute for Quantum Optics and Quantum Information, Austrian Academy of Sciences, Boltzmanngasse 3, Vienna, Austria}



\date{\today}

\begin{abstract}
Quantum field theory is the application of quantum physics to fields. It provides a theoretical
framework widely used in particle physics and condensed matter physics. One of the most distinct
features of quantum physics with respect to classical physics is entanglement or the existence of strong
correlations between subsystems that can even be spacelike separated. In quantum fields, observables
restricted to a region of space define a subsystem. While there are proofs on the existence of local
observables that would allow a violation of Bell’s inequalities in the vacuum states of quantum fields as
well as some explicit but technically demanding schemes requiring an extreme fine-tuning of the
interaction between the fields and detectors, an experimentally accessible entanglement witness for
quantum fields is still missing. Here we introduce smeared field operators which allow reducing the
vacuum to a system of two effective bosonic modes. The introduction of such collective observables is
motivated by the fact that no physical probe has access to fields in single spatial (mathematical) points but
rather smeared over finite volumes. We first give explicit collective observables whose correlations reveal
vacuum entanglement in the Klein-Gordon field. We then show that the critical distance between the two
regions of space above which two effective bosonic modes become separable is of the order of the
Compton wavelength of the particle corresponding to the massive Klein-Gordon field.
\end{abstract}
\pacs{03.70.+k, 03.67.Mn, 03.67.-a}
\maketitle

\section{Introduction}\label{sec: intro}
Entangled states of composed quantum systems are a subject of particular interest since they manifest genuinely quantum mechanical properties  $-$ they yield correlations between observables measured on the subsystems that cannot be explained by any local realistic model \cite{bell}. Since entanglement is the primary resource that allows quantum communication and computation protocols to outperform classical ones \cite{nielsen_chuang}, its investigation reaches far beyond fundamental concepts of quantum physics. It also has a central role in some macroscopic properties of solids - internal energy, heat capacity or magnetic susceptibility can reveal the existence of entanglement within solids in the thermodynamical limit \cite{wiesniak, brukner, toth}. Entanglement is further related to superfluidity, the Meissner effect and flux quantization in superconductors as well as long range order correlations in Bose-Einstein condensates (BEC) \cite{amico_fazio_et.al}.

A natural framework for considering systems composed of parts, which are associated with disconnected regions of space, is quantum field theory with its causal and local structure, where one may treat fields supported on spacelike separated regions as subsystems. Here we will consider the entanglement of the relativistic vacuum state. This is not only thought to be connected with black holes thermodynamics \cite{bombelli} or the holographic principle \cite{bousso}, but also to manifest itself in the Bekenstein-Hawking black hole radiation \cite{hawking} and Unruh acceleration effects \cite{Unruh}.

There are a number of studies proving entanglement between spatial regions in the ground state of relativistic quantum field theory. The central result underlying all the ``proofs in principle'' is the Reeh-Schlieder theorem \cite{reeh}, formulated in the language of algebraic quantum field theory. Let $A$ be a space-time region and ${\cal A}(A)$ the local algebra of all the operators with support in $A$
\footnote{More precisely, ${\cal A}(A)$ is defined as the algebra generated by operators of the form $\int_A dxdt\, f(\vec x,t)\hat\Phi(\vec x,t)$, $\int_A dxdt\, g(\vec x,t)\hat\pi(\vec x,t)$ with test functions $f$, $g$.}. The theorem states that, for any state of finite energy $|\psi\ket$ (in particular, the vacuum) the subspace ${\cal A}(A)|\psi\ket$ is dense in the entire Hilbert space, which means that for an arbitrary state $|\psi'\ket$, we can find a sequence of \textit{local} operations $\Pi_n$ such that $\lim_{n\rightarrow \infty}\Pi_n|\psi\ket=|\psi'\ket$. In particular, we can reconstruct an arbitrary state $|\psi_B\ket$ in region $B$ by applying such operations in $A$ and then tracing outside $B$. The vacuum is thus an entangled state as it allows for a remote state preparation \cite{bennet}.

As a consequence of the Reeh-Schlieder theorem and the positivity of a partially transposed separable density matrix, it is possible to prove that causally separated local regions are entangled in the vacuum state \cite{narnhofer}. There exist operators $\hat A$, $\hat B$ in algebras ${\cal A}(A)$, ${\cal B}(B)$ associated with causally separated regions $A$, $B$ that allow for constructing an entanglement witness (an operator whose mean value is larger or equal to zero for all separable states), which is violated (i.e. negative) in the vacuum reduced to these two regions. It is possible to choose operators $\hat A$, $\hat B$ as annihilation or creation operators of some local bosonic modes but the explicit construction is not known. 

It is also possible to prove the existence of local, bipartite observables that allow violation of Bell inequalities in the vacuum state \cite{werner}, but, again, the specific form of these observables is not known. Another result is that it is possible to locally couple two qubits (detectors) to the field in such a way that, after a finite time, the reduced state of such a pair of two-level systems has negative partial transpose \cite{reznik}. In this last case, one needs an explicit expression for the space-time dependence of the detector-field couplings. In order to prove entanglement for arbitrary separation of the detectors, these couplings need to be of a very specific, fine-tuned, form, involving superoscillating functions that require switching the sign of the interaction between the field and probes during the experiment. This seems technically extremely demanding. Along different lines (without referring to observables) in \cite{plenio} entanglement between two separated segments of one-dimensional free Klein-Gordon field in the vacuum state is quantified by the logarithmic negativity which is investigated numerically and shown to be finite for both critical and noncritical field limit.

To which extent can the vacuum of a quantum field be operationally accessed and used as an entanglement resource? By integrating over scalar Klein-Gordon field operators with compactly supported real functions  -- i.e.\ detection profiles localized in two regions of space -- we define collective field operators (and collective conjugate momenta). These weighted averages of operators allow reducing the vacuum to a system of two effective bosonic modes. To such a reduced state we apply the entanglement measure for continuous variables systems based on Simon's criterion for separability \cite{simon}. This approach has several advantages. Since separability criteria for infinite-mode states are unknown, we need to reduce the vacuum to a more comprehensible system. Our approach enables to quantify entanglement present in the resulting two-mode system, which is still infinite dimensional, unlike in  \cite{reznik}, where entanglement is first transferred to two qubits.  Physical probes have finite spatial resolution, so introducing collective observables is a reasonable first approximation towards a more realistic treatment of the problem. Finally, the Klein-Gordon field is the continuum limit of an infinite linear harmonic chain, and within this ``collective approach'' entanglement between blocks of oscillators in the ground state of the chain was found and quantified in \cite{kofler}. It is therefore interesting in itself to understand the relation between the field and the chain from this particular perspective.

In this paper we prove the existence of a critical distance between two regions of space above which two effective bosonic modes associated with the regions become separable. From the numerical analysis of the linear harmonic chain this critical distance is estimated to be of the order of the Compton wavelength of the particle corresponding to the massive one-dimenisonal Klein-Gordon field (the continuum limit of the linear harmonic chain). We also give an explicit example of the possible profiles that allow for a construction of entangled modes. Numerical results obtained for this exemplary functions are presented and discussed.

The structure of the paper is as follows: In Sec.\ \ref{sec:separability} the entanglement criterion and measure, which are further applied, are introduced and commented. Section \ref{sec:collective} is devoted to the collective operators in the relativistic scalar quantum field theory framework. We give their definition in terms of the detection profiles and discuss the constraints on the latter. Next, we derive one of the main results: the proof of the existence of a critical distance above which the modes defined within the collective operators approach, become separable. The section is closed with an explicit construction of the interaction needed to read out our observables. Section \ref{sec:discrete} deals with the linear harmonic chain. First, the relation between harmonic chain and continuous field is briefly reviewed. Second, on the basis of numerical analysis, the optimal profiles (maximizing the entanglement witness) in the discrete case are described and the critical distance for them is obtained. For the optimal profiles in the continuum limit this distance is then estimated to be of the order of the particles' Compton wavelength. In Sec.\ \ref{sec:field} we give an explicit example of profiles that allow for defining entangled modes and present the numerical results obtained for this special case. We close the paper with final remarks and conclusion in Sec.\ \ref{sec:conclusion}.

\section{Separability criteria}\label{sec:separability}
In this paper we use a particular form of the separability criterion derived by Simon \cite{simon}, necessary and
sufficient for two-mode Gaussian states. Following the original notation, we introduce a vector of phase space operators for the two modes system:
\begin{displaymath}
\hat \xi \equiv (\hat Q_A, \hat P_A, \hat Q_B, \hat P_B)^T\,
\end{displaymath}

where $T$ denotes transposition. Canonical commutation relations (CCR) in natural units, i.e.\ $\hbar=c=1$, can be concisely written in a matrix form
\begin{displaymath}
[\hat\xi_i,\hat\xi_j]=i\Omega _{ij},
\end{displaymath}
where we use the two-mode symplectic matrix
\begin{displaymath}
\Omega:=\left(
\begin{array}{cc}
 0 & 1  \\
-1 & 0 
\end{array}
\right)\bigoplus
\left(
\begin{array}{cc}
 0 & 1  \\
-1 & 0 
\end{array}
\right).
\end{displaymath}
 
Defining the variance matrix $V_{ij}:=\frac{1}{2}\left\langle \{\hat\xi_i - \langle \hat\xi_i\rangle, \hat\xi_j -\langle \hat\xi_j\rangle \} \right\rangle$, where $\{\hat A, \hat B\}:=\hat A \cdot \hat B + \hat B\cdot \hat A$, we obtain a compact statement of the Heisenberg uncertainty relations:
\begin{equation}\label{vmc}
V+\frac{i}{2}\Omega\geq 0.
\end{equation}
Every physical state has to satisfy this inequality. For separable states it must hold also after partial transposition \footnote{For relativistic quantum field theory partial transposition on the level of field operators is a partial CPT (completely positive and trace preserving)-map combined with reflection \cite{narnhofer}. In our case it reduces to partial time reversal, exactly as in nonrelativistic continuous variables systems \cite{simon}.}. The effect of partial transposition at the level of variance matrix elements is only $\langle \hat P_A\hat P_B\rangle \rightarrow -\langle \hat P_A\hat P_B\rangle$, provided that the system also satisfies
\begin{equation}\nonumber
\begin{array}{ll}
   \mbox{A1.} & \langle \hat Q_i\rangle=\langle \hat P_i\rangle=0\, ,\;\;\; i\in\{A,B\}\\
   \mbox{A2.} & \frac{1}{2}\langle \{ \hat Q_i,\hat P_j \}\rangle = 0\,,\;\;\; i,j\in\{A,B\}\,, \\
\end{array}
\end{equation}
which is always the case for us (one can always find a local symplectic transformation that enforces A1 and A2 \cite{simon, duan}).
Thus, all physical separable states fulfilling A1, A2, satisfy the following inequality (Simon's criterion):
\begin{equation}\label{critgen}
\begin{array}{c}
\frac{1}{4} - \langle{\hat Q_A}^2\rangle \langle{\hat P_A}^2\rangle - \langle{\hat Q_B}^2\rangle \langle{\hat P_B}^2\rangle - 2|\langle \hat Q_A\hat Q_B\rangle \langle \hat P_A\hat P_B\rangle | + \\
\\
+ 4 \left(\langle{\hat Q_A}^2\rangle\langle{\hat Q_B}^2\rangle - \langle \hat Q_A\hat Q_B\rangle^2\right)\left(\langle{\hat P_A}^2\rangle\langle{\hat P_B}^2\rangle - \langle \hat P_A\hat P_B\rangle^2\right) \geq 0    \;.
\end{array}
\end{equation}
In the case that the system additionally satisfies
\begin{equation}\label{ass}\nonumber
\begin{array}{ll}
 \mbox{A3.} & \langle{\hat Q_A}^2\rangle=\langle {\hat Q_B}^2\rangle\, ,\;\;\; \langle {\hat P_A}^2\rangle=\langle {\hat P_B}^2\rangle\,, \\
\end{array}
\end{equation}
we can considerably simplify the separability criterion by first factorizing it
\begin{equation}
\begin{array}{c}
\left\{\frac{1}{4}-\left(\langle {\hat Q_A}^2\rangle -|\langle\hat Q_A \hat Q_B\rangle|\right)\left(\langle {\hat P_A}^2\rangle - |\langle\hat P_A \hat P_B\rangle |\right)\right\}\!\times\nonumber \\
\\
\times \!\left\{\frac{1}{4} - \left(\langle {\hat Q_B}^2\rangle + |\langle\hat Q_A \hat Q_B\rangle |\right)\left(\langle {\hat P_B}^2\rangle+ |\langle\hat P_A \hat P_B\rangle |\right)\right\} \geq 0\,.
\end{array}
\end{equation}
Notice that the second factor is always non positive. If it equals zero, both correlations $\langle\hat Q_A \hat Q_B\rangle$ and $\langle\hat P_A \hat P_B\rangle$ must vanish, but then also the first factor equals zero. So finally, the above inequality is equivalent to
\begin{equation}\label{crit}
\left(\langle {\hat Q_A}^2\rangle -|\langle\hat Q_A \hat Q_B\rangle|\right)\!\cdot\! \left(\langle {\hat P_A}^2\rangle - |\langle\hat P_A \hat P_B\rangle |\right)\geq \frac{1}{4}\;.    
\end{equation}
Condition A3 is not satisfied for general profiles, but can be justified by physical assumptions (e.g., using the same detectors in regions A and B). In Sec.\ \ref{sec:discrete} this assumption will also be supported by numerical results. We also have some evidence that profiles that minimize the left-hand side of \eqref{critgen} satisfy A3 (see Sec.\ \ref{sec:discrete}).

To quantify entanglement, we will use the degree of entanglement given by
\begin{equation}\label{epsilon}
\varepsilon =1-4\left(\langle {\hat Q_A}^2\rangle -|\langle\hat Q_A \hat Q_B\rangle|\right)\!\cdot\! \left(\langle {\hat P_A}^2\rangle - |\langle\hat P_A \hat P_B\rangle |\right)\;.
\end{equation}
For Gaussian states $\varepsilon > 0$ iff the state is entangled. In this case $\varepsilon$ is a monotonically increasing function of the negativity $\mathcal{N}$ (the absolute sum of the negative eigenvalues of the partially transposed density matrix): $\varepsilon = \frac{\mathcal{N}}{\mathcal{N}+\frac{1}{2}}$. The negativity is based on the Peres-Horodecki criterion \cite{peres, horodecki} and was shown to be an entanglement monotone \cite{negativity1, negativity2}.

In the literature there are plenty of other criteria which, as well as the one we use, are necessary and sufficient for Gaussian states and also are phrased with second order correlations between phase space operators (e.g.\ \cite{duan}, \cite{kim}, \cite{anders}). It is then obvious that for Gaussian states they all yield the same results. However, for non-Gaussian states those criteria are only sufficient, so we should ask which of these is in general the strongest. In other words: are there any non-Gaussian entangled states such that one criterion out of those mentioned above would detect it, whereas some other would fail? The answer is negative: either all of them will be satisfied or all violated. From a physical point of view this is clear, simply because we cannot discriminate Gaussian from non-Gaussian states on the basis of their variance matrices (second order correlations). Mathematically, it can be shown that all those criteria are given by the same inequality up to a local linear canonical transformation of modes. This means that the choice of any of these criteria results in the same set of states detected as entangled. In any case, as we are going to see, in this work we will only be concerned with Gaussian states, so that \eqref{critgen} (and \eqref{crit}, when A3 is met) will always provide necessary and sufficient conditions for separability.

\section{Collective operators}\label{sec:collective}

We consider a massive Klein-Gordon (KG) field. The field $\hat\phi(\vec x, t)$ and conjugate momentum $\hat\pi(\vec x ,t)$ satisfy the equal-time canonical commutation relations, \cite{stone} 
\begin{equation}\label{eqtimecr}
\begin{array}{l}
\displaystyle [\hat \pi (\vec x ,t),\,\hat \phi (\vec y, t)]=i\,\delta^3(\vec x-\vec y)\\
\displaystyle [\hat \phi (\vec x ,t),\,\hat \phi (\vec y, t)]=0\\
\displaystyle [\hat \pi (\vec x ,t),\,\hat \pi (\vec y, t)]=0\\
\end{array}
\end{equation}

They can be expanded in terms of creation and annihilation operators, $\hat a_{\vec k}$ and $\hat a_{\vec k}^\dag$ of normal modes:
\begin{eqnarray}
\hat \phi (\vec x,t)=\frac{1}{(2\pi)^{3/2}}\int_{-\infty}^{+\infty}d^3 k\,\sqrt{\frac{1}{2\varpi_k}} \left( \hat a_{\vec k} e^{i\vec k\vec x -i{\varpi}_k t} + {\hat a_{\vec k}}^\dag e^{-i\vec k\vec x + i{\varpi}_k t} \right)\nonumber\\
\hat \pi (\vec x,t)=\frac{-i}{(2\pi)^{3/2}}\int_{-\infty}^{+\infty}d^3 k\, \sqrt{\frac{\varpi_k}{2}} \left( \hat a_{\vec k} e^{i\vec k\vec x -i{\varpi}_k t} - {\hat a_{\vec k}}^\dagger e^{-i\vec k\vec x + i{\varpi}_k t} \right)\;,\nonumber
\end{eqnarray}
where ${\varpi}_k =\sqrt{\vec k ^2+m^2}$.

The vacuum state is defined by the property
\begin{equation}\label{vacuum}
     \hat a_{\vec k}|0\ket=0 \;\; \forall \vec k\;.
\end{equation}

We study the possibility to detect entanglement in the vacuum state when the allowed measurements are constrained to field operators \textit{smeared} over two bounded space-time regions (\textit{collective} field observables). Our motivation is twofold: first, a field operator in a single space-time point is not a physical observable but a purely mathematical concept. Physical probes always have finite spatial resolution, so introducing collective observables is a reasonable first approximation towards a more realistic treatment of the problem. Second, since separability criteria for infinite-mode states are unknown, we want to reduce the vacuum to a system consisting of only two bosonic modes. In a general case, the smearing is given by two different real functions $g_A(\vec x)$, $g_B(\vec x)$ with compact supports and collective operators in the regions $A$, $B$ are defined as follows:
\begin{equation}\label{coll}
\begin{array}{c}
\displaystyle \hat Q(\vec x_{A/B})\equiv \hat Q_{A/B}:=\int_{-\infty}^{+\infty}d^3x\, g_{A/B}(\vec x-\vec x_{A/B})\hat \phi (\vec x, t)\\
\displaystyle \hat P(\vec x_{A/B})\equiv \hat P_{A/B}:=\int_{-\infty}^{+\infty}d^3x\, g_{A/B}(\vec x-\vec x_{A/B})\hat \pi (\vec x, t)\;.
\end{array}
\end{equation}
We consider collective operators that satisfy CCR
\begin{equation}\label{CCR}
\begin{array}{l}
\displaystyle [ \hat Q_{A},\,\hat P_{A}]=i \;,\\
\displaystyle [ \hat Q_{A},\,\hat P_{B}]=0 \;,\\
etc\,...
\end{array}
\end{equation}
which hold if the profiles satisfy orthonormality conditions (to be discussed in the next paragraph). The fact that collective phase space operators satisfy CCR guarantees that indeed we deal with two distinct bosonic modes, so it is meaningful to treat them as subsystems and speak about entanglement or separability of their joint state\footnote{More precisely, the collective operators on $A$ and $B$ generate two commuting subalgebras, that in turn induce two subsystems in the Hilbert space \cite{zan}, each of which is isomorphic to the space of a one-dimensional particle.}.

Completing the set $\{g_{A}(\vec x - \vec x_A), g_B (\vec x - \vec x_B)\}$ to an orthonormal basis in $L^2(\mathbb {R}^3)$, (\ref{coll}) can be extended to a linear canonical transformation of modes, two of which coincide with our collective ones. Tracing the global ground state over all but these particular two modes gives the final state of the two subsystems. Since the vacuum of a scalar quantum field is Gaussian and both transformations preserve this property, the final reduced state is Gaussian as well. This observation is very important as it means that the criterion \eqref{crit} is in our case necessary and sufficient (if condition A3 is satisfied) and \eqref{epsilon} indeed gives the degree of entanglement between the two bosonic modes. It is clear that the expectation values of all the possible products and combinations of the collective operators \eqref{coll} in the global vacuum \eqref{vacuum} coincide with their values calculated with respect to the reduced state.

The idea of restricting the possible measurements to a pair of collective modes is an extension of \cite{kofler} (where the chain of harmonic oscillators is considered) to the framework of scalar quantum field theory (QFT) with general profiles. Our main goal is to explicitly find two profiles $g_{A/B}(\vec x)$ such that the effective modes \eqref{coll} are entangled.

\subparagraph{Conditions on the profiles.}
Before proceeding to prove our main results, we need explicit expressions for the constraints that the detection profiles $g_{A/B}(\vec x)$ have to satisfy. It is useful to express them in terms of the Fourier transform
\begin{displaymath}
g(\vec k):= \frac{1}{(2\pi)^{3/2}}\int_{-\infty} ^ {+\infty} d^3 x\, e^{-i\vec k\vec x}g(\vec x)\,.
\end{displaymath}
We require that the collective operators \eqref{coll} satisfy CCR \eqref{CCR}. All the relations involving only collective position or only momentum operators are automatically satisfied due to \eqref{eqtimecr}. The remaining ones lead to orthonormalization of $\{g_{A}(\vec x - \vec x_{A}), g_{B}(\vec x - \vec x_{B})\}$ in  $L^2(\mathbb{R} ^3)$.
With $i,j=A,B$, we have
\begin{equation}\label{comutator} [\hat Q_{i}, \hat P_{j}]=i\,\delta _{i j} \iff \int_{-\infty}^{+\infty}d^3k\, e^{-i\vec k(\vec x_i-\vec x_j)} g_{i}(\vec k) g_{j}(-\vec k)=\delta _{i j}
\end{equation}
Further, all correlations in (\ref{crit}) should be finite. From the Cauchy-Schawrz inequality $|\langle{\hat Q_A \hat Q_B}\rangle|\leq \sqrt{\langle{\hat Q_A}^2\rangle \langle{\hat Q_B}^2\rangle}$ and $|\langle{\hat P_A \hat P_B}\rangle|\leq \sqrt{\langle{\hat P_A}^2\rangle \langle{\hat P_B}^2\rangle}$. So, (with assumption A3) it is only necessary that $\langle{\hat Q_A}^2\rangle$ and $\langle{\hat P_A}^2\rangle$ are finite. This indeed holds, if the Fourier transforms of the smearing functions decay fast enough\footnote{This condition was not satisfied in a paragraph devoted to scalar quantum field in \cite{kofler}. However, this fact does not affect the validity of the results obtained there for the linear harmonic chain}:
\begin{equation}\label{decay}
 |\hat g_{A/B}(\vec k)| \leq  \frac{1}{{|\vec k |}^{\lambda}} \, \;\;\mbox{for}\,|\vec k| \to {\infty}\,,\;\;\;\lambda > \frac{d+1}{2}\;\;\;\;\;\mbox{in $d$ space dimensions}.
\end{equation}
Finally, we demand A1-A3. Note that A1, A2 are satisfied for every profile, while A3 is equivalent to the additional conditions
\begin{eqnarray}
\langle {\hat Q_{A}}^2 \rangle = \langle {\hat Q_{B}}^2 \rangle &\iff& \int_{-\infty}^{+\infty}d^3k\,\frac{1}{2\varpi_k}| g_{A}(\vec k)| ^2 = \int_{-\infty}^{+\infty}d^3k\,\frac{1}{2\varpi_k}| g_{B}(\vec k)| ^2\nonumber\\
\langle {\hat P_{A}}^2 \rangle = \langle {\hat P_{B}}^2 \rangle &\iff& \int_{-\infty}^{+\infty}d^3k\,\frac{\varpi_k}{2} |g_{A}(\vec k)| ^2 = \int_{-\infty}^{+\infty}d^3k\,\frac{\varpi_k}{2} |g_{B}(\vec k)| ^2 \nonumber\,.
\end{eqnarray}

All correlations appearing in the criterion \eqref{crit} in terms of the Fourier transforms of the profiles read ($i,j=A,B$)
\begin{equation}\label{cor}
\begin{array}{ccc}
\displaystyle\langle \hat Q_i\hat Q_j \rangle &\displaystyle = &\displaystyle\int_{-\infty}^{+\infty}d^3k\, \frac{1}{2\varpi_k} e^{-i\vec k(\vec x_i - \vec x_j)} g_{i}(\vec k) g_{j}(-\vec k)\,,\\
\displaystyle\langle \hat P_i\hat P_j \rangle &\displaystyle = &\displaystyle\int_{-\infty}^{+\infty}d^3k\, \frac{\varpi_k}{2} e^{-i\vec k(\vec x_i - \vec x_j)} g_{i}(\vec k) g_{j}(-\vec k)\,.\\
\end{array}
\end{equation}
We will further denote $\vec D:=\vec x_B-\vec x_A$.  This parameter, appearing in $\langle\hat Q_A\hat Q_B\rangle$ and $\langle\hat P_A\hat P_B\rangle$, is a measure of the distance (separation) between the subsystems.

\subparagraph{Large separations limit.} It is natural to ask whether there exist profiles satisfying all the constraints given above and defining entangled modes for arbitrary separations. In this paragraph we will show that this is not possible.
Below we prove that for every pair of allowed profiles there exists a finite critical distance $D_{crit}$ such that corresponding collective operators are separable at distances larger than the critical one. More precisely: for every pair of orthonormal functions $g_{A/B}(\vec x)$, satisfying A3 and \eqref{decay} [giving finite correlations in \eqref{crit}] there exists $D_{crit}<\infty$ such that the modes defined by $g_A(\vec x - \vec x_A), g_B(\vec x - \vec x_B)$ are separable for every $|\vec D|>D_{crit}$. We will give the proof in three space dimensions but it remains valid in arbitrary finite dimensions.

First, notice that the integral form of the Cauchy-Schwarz inequality (i.e.\ H\"{o}lder's inequality) together with condition \eqref{decay}, not only guarantees the finiteness of all the correlations \eqref{cor} but, also asserts that their integrands are functions from $L^1(\mathbb{R}^3)$ (space of functions, which absolute value is integrable). In particular, it enables to prove that both $\frac{g_{A}(\vec k) g_{B}(-\vec k)}{\varpi_k} $ and $\varpi_k g_{A}(\vec k) g_{B}(-\vec k)$  are in $L^1(\mathbb{R}^3)$.

From Riemann-Lebesgue lemma \cite{simonreed} (it says that the Fourier transform of an $L^1$ function vanishes at infinity) it now immediately follows that
\begin{displaymath}
\lim_{|\vec D|\to\infty}\,\langle\hat Q_A\hat Q_B\rangle = \lim_{|\vec D|\to\infty}\,\int_{-\infty}^{+\infty}d^3k\, \frac{1}{2\varpi_k} e^{-i\vec k\vec D} g_{A}(\vec k) g_{B}(-\vec k) = 0,
\end{displaymath}
\begin{displaymath}
\lim_{|\vec D|\to\infty}\,\langle\hat P_A\hat P_B\rangle = \lim_{|\vec D|\to\infty}\,\int_{-\infty}^{+\infty}d^3k\, \frac{\varpi_k}{2} e^{-i\vec k\vec D} g_{A}(\vec k) g_{B}(-\vec k) = 0.
\end{displaymath}
In consequence, for $|\vec D|\rightarrow \infty$ the left-hand side of the criterion \eqref{crit} reduces to the product $\langle\hat Q_{A} ^2\rangle \langle\hat P_{A}^2\rangle$. Because of CCR imposed on the collective operators \eqref{CCR}, the Heisenberg uncertainty relation guarantees that this product is always greater or equal $\frac{1}{4}$, so in the limit of infinite separation the modes become separable. However, we obtain much stronger result by utilizing Eq.\ \eqref{cor}. Let us write the product $\langle\hat Q_{A} ^2\rangle \langle\hat P_{A}^2\rangle$ as a double integral symmetrized over the integration variables
\begin{displaymath}
\langle\hat Q_{A}^2\rangle \langle\hat P_{A}^2\rangle = \frac{1}{4}\iint_{-\infty}^{+\infty}d^3k \,d^3q\,\frac{1}{2}\left(\frac{\varpi_q}{\varpi_k}+\frac{\varpi_k}{\varpi_q}\right)\,|g_{A}(\vec k)\,g_{A}(\vec q)| ^2 \,>\frac{1}{4}\,.
\end{displaymath}
The last inequality is a direct consequence of the normalization of the profiles [conditions \eqref{CCR}], the fact that $\left(\frac{\varpi_q}{\varpi_k}+\frac{\varpi_k}{\varpi_q}\right)\geq 2$ for all $\vec k, \vec q \in \mathbb{R}^3$ and that the latter saturates only on the hyperplane $|\vec k|=|\vec q|$, which has zero Lebesgue measure.

Summarizing, we have shown that $\lim_{|\vec D|\to\infty}\,(\langle {\hat Q_A}^2\rangle - |\langle \hat Q_A\hat Q_B\rangle|)\!\cdot\! (\langle {\hat P_A}^2\rangle - |\langle \hat P_A\hat P_B\rangle|)> \frac{1}{4}$, which is equivalent to
\begin{equation}\label{limit}
\exists D_{crit}<\infty\; \mbox{such that} \;\left(\langle {\hat Q_A}^2\rangle -|\langle \hat Q_A\hat Q_B\rangle|\right)\!\cdot\! \left(\langle {\hat P_A}^2\rangle - |\langle \hat P_A\hat P_B\rangle |\right)\geq\frac{1}{4}\;\,\forall |\vec D| > D_{crit}\,. 
\end{equation}
This is, however, exactly the separability condition \eqref{crit}. So \eqref{limit} states that, given a pair of orthonormal functions $g_A(\vec x)$, $g_B(\vec x)$ which satisfy \eqref{decay} and A3, there exists $D_{crit}<\infty$ such that once $|\vec x_A - \vec x_B|\equiv |\vec D|>D_{crit}$, the modes defined by $g_A(\vec x - \vec x_A)$, $g_B(\vec x - \vec x_B)$ are separable.

It is interesting, although not very relevant from the physical point of view, that this result is not restricted to functions with compact support, as this property has no specific role in the proof (only orthonormality is needed). For example, it holds for any pair of orthonormal test functions (element of Schwartz space:\ rapidly decreasing functions) satisfying A3.

\subparagraph{Measurement of the collective observables. }\label{measure}
A significant difference between our and other related works lies in the treatment of the detector. We define it only in terms of the observables that it measures. In this paragraph we propose a model by which a system interacting with the field can be used to implement the desired measurements of the collective observables. We will follow the method discussed in \cite{peresbook}. In order to measure each of the operators we thus need a different interaction. In general, however, to either of the effective modes $A, B$ we couple another bosonic mode -- a detector. Then, under the coupling of the suitable degrees of freedom of the subsystem and the detector, measurements made on the latter reveal the value of the corresponding observable.

Let us consider a universal situation when observable $\hat W$ is measured on mode $i$, with $i={A, B}$. The Hilbert space $\mathcal{H}_{d i}$ of the joint system comprised of the subsystem $i$ (with Hilbert space denoted by $\mathcal{H}_i$) and its detector (with Hilbert space $\mathcal{H}_d$) is a tensor product $\mathcal{H}_{d i}=\mathcal{H}_i\otimes\mathcal{H}_d$. We introduce the phase space observables for the detector $\hat Q_d, \hat P_d$, which satisfy canonical commutation relations $[\hat Q_d, \hat P_d]=i$. In other words, we consider the detector's state space to be isomorphic to the Hilbert space of a one-dimensional particle. An interaction Hamiltonian, which allows a  measurement of the operator $\hat W$, takes the form
\begin{equation}\label{interact}
\hat H_I^{W}:=\alpha \hat W \hat P_d \,,
\end{equation}      
where $\alpha$ is a time independent coupling constant for this particular interaction. As bases of the Hilbert spaces $\mathcal{H}_i, \mathcal{H}_d$ we take the sets of eigenvectors of $\hat W, \hat Q_d$, namely $\{|w\ket\}_{w\in \mathbb{R}}, \{|q_d\ket\}_{q_d\in \mathbb{R}}$ such that $\hat W|w\ket =w|w\ket$ and $\hat Q_d|q_d\ket =q_d|q_d\ket$. (We consider here observable $\hat W$ with continuous spectrum because we are primary interested in measuring the collective observables, however the framework considered below applies also to operators with discrete spectrum.) The state of the subsystem $i$ is mixed. In the chosen basis it can be written as
\begin{equation}\label{stateA}
\hat \rho_i = \iint_{-\infty}^{+\infty}dw\,dz\, \rho_i (w, z)|w\ket\bra z|
\end{equation}
The initial state of the detector is, in an idealized situation, a pure eigenstate $|q_d\ket$ of $\hat Q_d$. Therefore, prior to the interaction, the joint state $\hat \rho_0$ of the two systems is
\begin{equation}
\hat \rho_0 = \iint_{-\infty}^{+\infty}dw\,dz\, \rho_i (w, z)|w\ket\bra z|\otimes |q_d\ket\bra q_d|\,.  
\end{equation}
If the time scale of the measurement process is much smaller then that of the free evolution of the field and the detector, the time evolution of the density matrix $\hat \rho_0$ is given by the Hamiltonian \eqref{interact}:
\begin{equation}
\hat \rho_t = e^{-i \hat{H}_I^W t}\hat \rho_0 e^{i \hat{H}_I^W t}.
\end{equation}
Making use of the fact that the momentum operator is a generator of spatial translations (see e.g. \cite{ll}) we notice that
\begin{equation}
e^{-i \hat{H}_I^W t}|w\ket\otimes |q_d\ket = |w\ket\otimes|q_d + \alpha t w\ket\, ,
\end{equation}
so finally
\begin{equation}
\hat \rho_t = \left(\frac{1}{\alpha t}\right)^2\int_{-\infty}^{+\infty}dw\,dz\, \rho_i \left(\frac{w- q_d}{\alpha t}, \frac{z-q_d}{\alpha t}\right)\left|\frac{w- q_d}{\alpha t}\right\rangle\left\langle \frac{z-q_d}{\alpha t}\right|\otimes |w\ket\bra z|\,.
\end{equation}
Performing measurements on the detector's degrees of freedom, we can reconstruct the values of the observable $\hat W$ in the state of subsystem $i$ given by \eqref{stateA}. It is straightforward to derive that the probability amplitude for the detector to be at time $t$ in some eigenstate $|a\ket$ of $\hat Q_d$ is proportional to $\rho_i\left(\frac{a-q_d}{\alpha t},\frac{a-q_d}{\alpha t}\right)$, which is in turn the probability amplitude of subsystem $i$ being in state $\left|\frac{a-q_d}{\alpha t} \right\rangle$. A more realistic treatment would involve assuming for the initial state of the detector not an eigenstate of $\hat Q_d$ but rather a superposition $\int_{-\infty}^{+\infty}dq_d f(q_d)|q_d\ket$ [e.g.\ for a coherent initial state $f(q)$ would be a Gaussian packet]. In such a case the amplitude for the detector to be at time $t$ in the state $|a\ket$ is proportional to $\int_{-\infty}^{+\infty}dq_d\, \rho_i(\frac{a-q_d}{\alpha t},\frac{a-q_d}{\alpha t})\left|f(q_d)\right|^2$.

In the above example the considered subsystem was coupled to the detector's degree of freedom which corresponds to the operator $\hat P_d$. This choice is of course arbitrary, i.e.\ equivalently well the other degree of freedom, corresponding to $\hat Q_d$, may be utilized to perform the measurement. In such a case the interaction takes form $\hat H_I^{\prime W}:=\beta \hat W \hat Q_d $, where $\beta$ is again a time independent coupling constant. As a basis of the Hilbert space $\mathcal{H}_d$ we take the set of eigenvectors of $\hat P_d$ and, as the initial state of the detector, we consider the eigenstate of $\hat P_d$. Following all the previous steps with these changes in mind, we obtain that measuring this detector's state in the momentum basis, again, enables to reconstruct the value of the observable $\hat W$ on the state of subsystem $i$. 

Although we couple each mode with an (effective) one-dimensional particle, different degrees of freedom of the measured modes are involved in measurements of conjugate collective observables. So, in the outlined scheme the interactions are relatively simple but the detectors serve solely as devices to reconstruct the values of collective operators in the two-mode state considered. On the other hand, if the vacuum entanglement is actually transferred to the detectors (as in \cite{reznik}), by the price of a very fine tuned, time dependent interaction, it is possible to detect entanglement for arbitrary separations\footnote{If the detector model is described by a natural interaction, it is not possible to detect entanglement with its use \cite{fabio}.}.

\section{Results from discrete systems}\label{sec:discrete}

The result of the previous section says that no matter what are the shapes of the detection profiles, collective operators too distant from each other must be separable, but it does not give any indication about the possibility of finding entanglement by measurements in regions sufficiently close together. It is not trivial to tackle this problem directly, as the possible detection profiles form an infinite dimensional space and many of them will still define separable modes; it is therefore instructive to study a discretized version of the system, where numerical analysis can be performed. The numerics will give important insights into the shape of the profiles that can show entanglement and, furthermore, it will give evidence of the existence of a critical distance independent of the specific profile, thus strengthening the results of Sec.\ \ref{sec:collective}, and provide an estimation for it. Here only the one-dimensional case will be considered.
\subparagraph{Continuum limit.}
A one-dimensional bosonic field can be formally obtained as the continuum limit of a chain of coupled harmonic oscillators. This is especially useful for us, as the discrete system can be better analyzed; in particular, the freedom in the choice of profiles reduces to an optimization problem of functions of a finite number of degrees of freedom.

Let us briefly review the relation between the harmonic chain and the continuous field (see, e.g., \cite{botero}). The KG Hamiltonian
\begin{equation}
        \label{kg}
   \hat H_{KG}=\frac{1}{2}    \int dx\left(m^2\hat\phi(x)^2+ \hat\pi(x)^2 + (\nabla\hat\phi(x))^2\right)\,
\end{equation}
can be written as the limit for $\Delta x \rightarrow 0$ of  
\begin{eqnarray}\nonumber
  \hat H_{dis}&=&\frac{1}{2}    \Delta x \sum_{j}\left(m^2\hat\phi_j^2 + \hat\pi_j^2 + \frac{1}{\Delta x ^2} (\hat\phi_j-\hat\phi_{j-1})^2\right)\\
  &=&\frac{1}{2}    \sum_{j}\left(\Delta x m^2\hat\phi_j^2 + \Delta x\hat\pi_j^2 + \frac{1}{\Delta x} (\hat\phi_j-\hat\phi_{j-1})^2\right)\,,
  \label{discrete}
\end{eqnarray}
where the discretized field operators are defined as $\hat\phi_j:=\hat\phi(j\Delta x)$, $\hat\pi_j:=\hat\pi(j\Delta x)$, with $j$ integer and $\Delta x$ being the spacing between successive points.

The expression \eqref{discrete} can be put in correspondence with a chain of $N$ harmonically coupled oscillators, with conjugate observables satisfying $[\hat{\bar{q}}_j,\hat{\bar{p}}_k]=i\delta_{jk}$ and Hamiltonian
\begin{equation}\label{ham}
       \hat H= \frac{1}{2}\sum^{N}_{j=1}\left(M\omega^2\hat{\bar{q}}_j^2 + \frac{\hat{\bar{p}}_j^2}{M} + M\Omega^2\left(\hat{\bar{q}}_j-\hat{\bar{q}}_{j-1}\right)^2 \right)\,,
\end{equation}
where $M$ is the mass of each individual oscillator, $\omega$ its proper frequency and $\Omega$ the coupling frequency. Periodic boundary conditions $\hat{\bar{q}}_0=\hat{\bar{q}}_N$ are assumed.

If we set \eqref{discrete} equal to \eqref{ham}, we obtain
\begin{displaymath}
    \begin{array}{l}
     \Delta x m^2\hat\phi_j^2 = M\omega^2\hat{\bar{q}}_j^2 \,,\\
 \Delta x\hat\pi_j^2 = \frac{\hat{\bar{p}}_j^2}{M} \,, \\
 \frac{1}{\Delta x} (\hat\phi_j-\hat\phi_{j-1})^2 = M\Omega^2\left(\hat{\bar{q}}_j-\hat{\bar{q}}_{j-1}\right)^2 \,,
\end{array}
\end{displaymath}
from which we derive
\begin{equation}
\begin{array}{l}
    \hat\phi_j= \sqrt{\frac{M}{\Delta x}}\frac{\omega}{m}\hat{\bar{q}}_j\\
    \hat\pi_j=\sqrt{\frac{\Delta x}{M}}\hat{\bar{p}}_j\\
    m\Delta x = \frac{\omega}{\Omega}\;.
\end{array}
\label{contlim}
\end{equation}
In order to define a correct continuum limit, the scaling of the parameters in \eqref{ham} with $\Delta x$ must obey \eqref{contlim} with $m$ fixed.

In order to simplify the analysis, we can rewrite \eqref{ham} in the following form:
\begin{equation}\label{hamsimp}
 \hat H=\frac{E_0}{2} \sum^{N}_{j=1}\left(\hat q_j^2+\hat p_j^2-\alpha \hat q_j\hat q_{j-1}\right),
\end{equation}
where $E_0=\sqrt{2\Omega^2+\omega^2}$, $\alpha=2\Omega^2/(2\Omega^2+\omega^2)$ and we introduced the dimensionless variables $\hat q_j=C\hat{\bar{q}}_j$, $\hat p_j=\hat{\bar{p}}_j/C$, with $C=\sqrt{ M\omega(1 + 2
\Omega^2/\omega^2)^{1/2}}$. In this form, the system is characterized by a single dimensionless parameter, the coupling constant $\alpha$, which, by construction, is constrained to values $0<\alpha<1$. In this case, the continuum limit is obtained by setting
\begin{equation}
 \label{scaling}
 \alpha = \frac{1}{1 + \frac{1}{2} \Delta x^2 m^2}
\end{equation}
and letting $\Delta x \rightarrow 0$, with $m$ constant.

If we want to describe a region of size $L$ using \eqref{hamsimp} as a discrete version of a Klein-Gordon field, we have to consider in the chain a block with a number of sites
\begin{displaymath}
 n=\frac{L}{\Delta x}= Lm \sqrt{\frac{\alpha}{2-2\alpha}}\,.
\end{displaymath}
As $n$ has to be an integer, for some values of $\alpha$, $L$ and $m$ the expression above is not well defined. It can therefore be more convenient, especially for carrying out numerical computations, to express $\alpha$ as a function of $n$ and the physical length:
\begin{equation}
 \label{scaling2}
 \alpha = \frac{1}{1 + \frac{1}{2} \left(\frac{mL}{n}\right)^2}\,.
\end{equation}
This relation fixes the physical size of a region. By increasing $n$ and having the coupling constant scaling as in \eqref{scaling2}, one approaches in the limit a region of size $L$ of a KG field with mass $m$. It is worth stressing that it is not the number of points $n$ that determines the size of a region, rather, this needs to be fixed through the relation \eqref{scaling2}. Increasing $n$ only provides a more refined description of the system.
\subparagraph{Collective entanglement and optimal profiles.}
The discrete version of the collective operators \eqref{coll} can be defined for a block $A$ of $n$ sites in a chain:
\begin{equation}
\begin{array}{l}
\label{collective}
    \hat Q_A := \sum_{j=1}^n f_j \hat q_{j+l} \\
    \hat P_A := \sum_{j=1}^n f_j \hat p_{j+l}\,,
\end{array}
\end{equation}
where $l+1$ is the position of the first site of $A$ and $f_j$ is the detection profile, determining how much each site in the chain contributes to the collective observables (note that the indices of the profiles always run in the range $\left\{1,\dots,n\right\}$, regardless of the position of the block in the chain).  If the profile satisfies the normalization condition
\begin{equation} \label{norm}
\sum_{j=1}^n f_j^2 = 1\,,
\end{equation}
then the collective operators have canonical commutation relations $\left[\hat Q_A,\hat P_A\right]=i$ and the subsystem they define is a bosonic mode.

Two detectors placed in two regions $A$ and $B$, described by detection profiles $f_j$ and $g_j$ respectively, effectively detect two bosonic modes. If the global state is Gaussian (as is the case for the vacuum state) then also the reduced state over the two modes is so, as was discussed in Sec.\ \ref{sec:collective}; we can therefore apply Simon's criterion \eqref{critgen} to establish whether the two modes are entangled or not.
If $\varepsilon$ is the corresponding entanglement measure, $\varepsilon > 0$ meaning entanglement and $\varepsilon \leq 0$ separability, we can ask for which profiles $f_j$, $g_j$ $\varepsilon$ is maximized when all the parameters are fixed. If two profiles exist such that $\varepsilon_{max} > 0$, then we can conclude that it is in principle possible to see entanglement, while $\varepsilon_{max} \leq 0$ proves separability for all possible collective operators.

\begin{figure}[!ht]
\begin{center}
\includegraphics[trim = 0mm 14mm 0mm 14mm, clip]{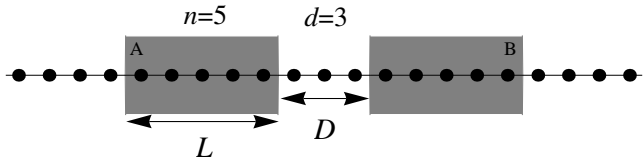}
\caption{Two blocks of a harmonic chain A and B. In this example, each block consists of $n=5$ oscillators and the blocks are separated by $d=3$ oscillators. The physical length $L$ of the blocks and their separation $D$ are related by $D=\frac{d}{n}L$}
\label{discrpar}
\end{center}
\end{figure}

Let us study systematically the case where the two regions $A$ and $B$ have the same length $L$. We want to know if, for a given separation $D$ between $A$ and $B$, it is possible in principle to find entanglement and what are the shapes of the profiles that maximize the entanglement measure. We will set from now on the mass of the field to $m=1$, so that all the lengths will be expressed in units of Compton wavelength $\lambda_c=\frac{1}{m}$. Note that $m$ is the only dimensional parameter in \eqref{kg}, so it defines the natural scale of the system. Using the discretized field, we can approach the problem numerically: we first fix the size $L$ of the regions, the number of sites (i.e.\ oscillators) in each block $n$ and the number of sites $d$ separating the two blocks. The physical separation $D$ is then given by\footnote{With the definition of the profiles that we use here, parameter $D$ defined in the last section gives exactly the distance between the near ends of the regions. This choice is obviously the most suitable for the analysis.}
\begin{equation}
\label{d}
D= \frac{d}{n} L\,
\end{equation}
(see Fig.\ \ref{discrpar}) and $\alpha$ is determined by the relation \eqref{scaling2}.

We consider a system of infinite total length, which means $\frac{N}{n} \gg 1$. We verified numerically that this limit is already well approximated for $N=10(2n+d)$. With all the parameters fixed, the entanglement measure $\varepsilon$ is a function of the $2n$ real variables $\left\{f_j,\,g_j\right\}_{j=1}^{n}$ and we can then find the numerical extremum with both $f$ and $g$ subject to the constraint \eqref{norm}. The first numerical evidence is that the optimal profiles are always mirroring, that is to say, $\varepsilon$ is always maximized by functions satisfying $g_j = f_{n+1-j}$; this allows us to reduce the problem to a maximization over $n$ variables. Furthermore, this symmetry ensures that $\langle  \hat Q_A^2 \rangle = \langle  \hat Q_B^2 \rangle$,  $\langle  \hat P_A^2 \rangle = \langle  \hat P_B^2 \rangle$ (condition A3 above), so that we can use the simplified entanglement measure \eqref{epsilon}.  Notice, that the obvious choice of a rectangular profile (a ``top-hat'' function) is far from optimal. Moreover, such a profile would not work in the continuum limit, as it gives diverging correlations.

We proceed in the following way: first we fix the physical region size $L$, then, for a given value of $d$, we look for the critical block size $n_{crit}(d)$ such that the blocks are entangled ($\varepsilon > 0$) for $n \geq n_{crit}(d)$ and separable ($\varepsilon \leq 0$) for $n<n_{crit}(d)$. Then we change $d$ and study the functional dependence between $n_{crit}$ and $d$; as $d\rightarrow\infty$ we approach the continuum limit. Assuming that $n_{crit}(d)$ is always finite (as it turns out, it is), we can expect three possible situations:

\begin{enumerate}
    \item \large$\frac{d}{n_{crit}(d)}\rightarrow \infty$ for $d\rightarrow \infty$.
   
    \normalsize This would mean that, in the limit, regions of size $L$ arbitrarily distant from each other can be entangled [remember that $L$ is the fixed physical length, while $D$ is given by \eqref{d}].
    \item \large $\frac{d}{n_{crit}(d)}\rightarrow 0$ for $d\rightarrow \infty$.
   
    \normalsize In this case, the physical distance $D$ below which we can see entanglement would vanish for the given region size $L$, so no entanglement could be seen between separated regions.
    \item \large $\frac{d}{n_{crit}(d)}\rightarrow $ $C(L)$ for $d\rightarrow \infty$, $0<C(L)<\infty$.
   
    \normalsize In this last case, there exists a distance $D(L) = C(L) L$ above which regions of size $L$ are always separable, but below which they can be entangled if the appropriate profile is chosen.
\end{enumerate}
As we are going to see, the data give strong numerical indication in favor of the third case.

Let us consider in detail, as an illustrating example, the results for $L=\sqrt{2}$ (the specific value is only chosen for numerical convenience). For fixed values of the separation between the blocks $d$ and of the number of sites in each block $n$, we search for the optimal profile that maximizes the entanglement parameter \eqref{epsilon}. We do so for $d$ fixed and increasing values of $n$, until we find $\varepsilon>0$. We repeat the procedure for $d=1,\dots,16$, so that finally we have, for each $d$, the smallest block size $n_{crit}(d)$ such that the parameter $\varepsilon$ (maximized over all profiles) is positive. The inverse coupling constant scales with $n$ as $\alpha^{-1} = 1 +  \frac{1}{n^2}$ [as required by \eqref{scaling2}]. The results are shown in Table \ref{table}.

\begin{table}[ht]
    \centering
    \begin{tabular}{|r|l|l|l|l|l|l|l|l|l|l|l|l|l|l|l|l|} \hline
            $d=$ & 1 & 2 &3&4&5&6&7&8&9&10&11&12&13&14&15&16  \\ \hline
            $n_{crit}(d)=$ & 2 &8 & 14& 20& 25& 31& 37& 43& 48& 54& 60& 65& 71& 77& 83&88 \\ \hline
    \end{tabular}
    \caption{Minimal block size $n_{crit}$ that allows entanglement for fixed physical size of the regions $L = \sqrt{2}$ as a function of the number $d$ of oscillators separating the two regions.}
    \label{table}
\end{table}

\begin{figure}[!ht]
\centering
\begin{minipage}[l]{0.33\linewidth}
\includegraphics[width=5.0cm]{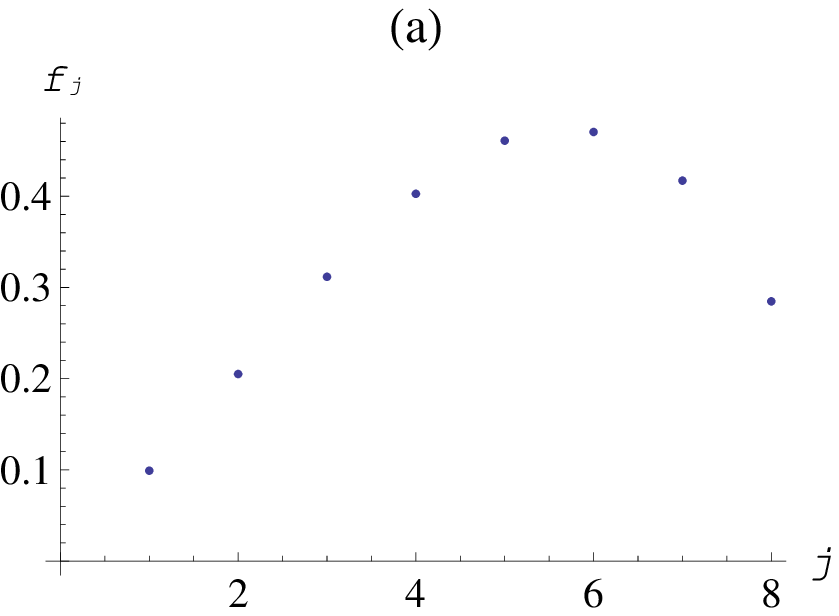}
\end{minipage}%
\begin{minipage}[c]{0.33\linewidth}
\includegraphics[width=5.0cm]{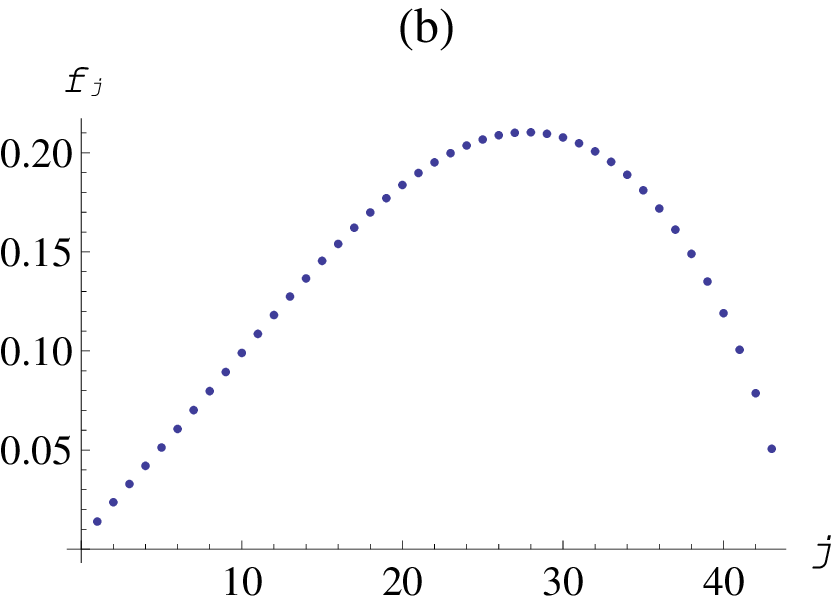}
\end{minipage}
\begin{minipage}[r]{0.33\linewidth}
\includegraphics[width=5.0cm]{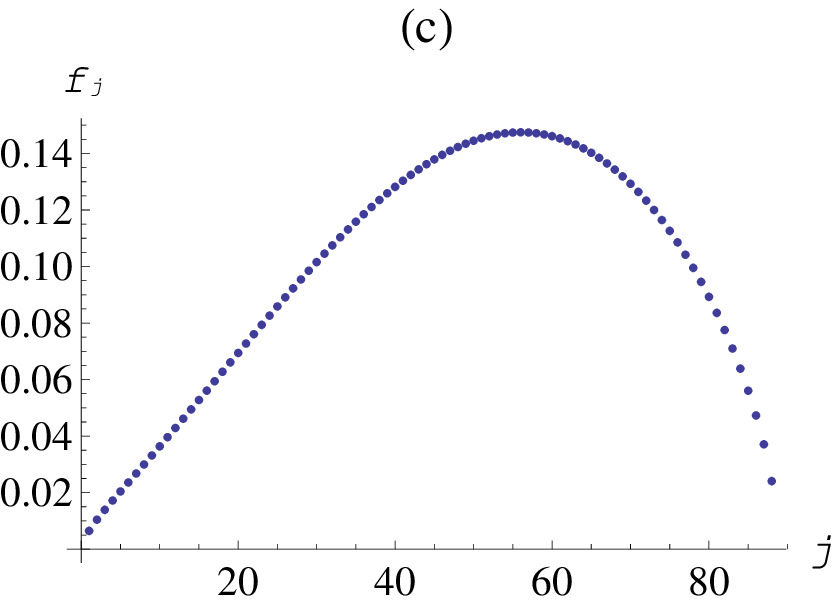}
\end{minipage}
\caption{Three optimal (maximizing the entanglement measure $\varepsilon$) profiles for fixed physical size of the regions $L = \sqrt{2}$. The number $n_{crit}$ of oscillators in one block is the smallest that allows entanglement for the given number $d$ of oscillators between the two blocks. For plot (a) $d=2$, $n_{crit}=8$, for (b) $d=8$, $n_{crit}=43$ and for (c) $d=16$, $n_{crit}=88$. The plotted profiles are for the left block; the profiles for the block on the right have the same shape, mirrored.}
\label{profiles}
\end{figure}

One can see that the relation between $n_{crit}$ and $d$ is approximately linear, corroborating hypothesis 3 above. In Fig.\ \ref{profiles} three optimal profiles are plotted, from which it can be clearly seen how the same shape is reproduced while increasing the number of points, as it is expected when approaching the continuum limit. We can conjecture that the continuous curve appearing in the limit $n\rightarrow \infty $ would correspond to the optimal profile in the continuum.

It is possible to extract the limiting value $C(L)=\mbox{lim}_{d\rightarrow\infty} \frac{d}{n_{crit}(d)}$ by linear interpolation of the data in Table \ref{table}. We obtain $C(L=\sqrt{2})=0.17$. From this, using relation \eqref{d}, we can calculate the physical distance below which regions of size $\sqrt{2}$ can be entangled: $D=C(L) L = 0.25$ (expressed in units of the Compton wavelength).

\begin{figure}[!ht]
\centering
\begin{minipage}[c]{0.4\linewidth}
\includegraphics[width=5.5cm]{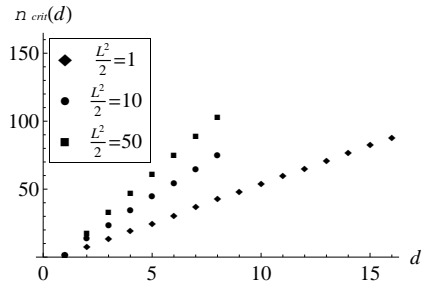}
\end{minipage}
\caption{Dependence of the critical number $n_{crit}(d)$ of sites within the blocks on the number $d$ of sites between the blocks for different region sizes $L$. The critical value $n_{crit}(d)$ is defined as the minimal number of sites that gives entanglement of the corresponding collective operators. For each $L$ a linear law is evident, the linear coefficient is defined as $1/C(L)$.}
\label{fig:linear}
\end{figure}

The same calculation can be repeated for different region sizes $L$. The linear behavior is confirmed for all the cases, as can be seen in the examples plotted in Fig.\ \ref{fig:linear}.

Once the linear dependence of $n_{crit}$ on $d$ is established, it is possible to calculate for each value of $L$ the coefficient $C(L)$.
\begin{figure}[!ht]
\centering
\begin{minipage}[c]{0.4\linewidth}
\includegraphics[width=5.5cm]{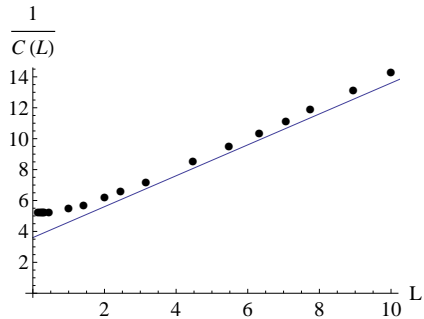}
\end{minipage}
\caption{Dependence of the coefficient $C(L)$ on the region size $L$. In the graph $\frac{1}{C(L)}$ is compared with a linear law.}
\label{fig:C}
\end{figure}
As can be seen from Fig.\ \ref{fig:C}, it is reasonable to infer that $\frac{1}{C(L)}$ has an asymptote for $L\rightarrow \infty$, namely $C(L)\sim \frac{c}{L}$; in the figure the plot of $\frac{1}{C(L)}$ is compared with a handmade asymptote with $c=1$. This result implies that, as the region sizes become arbitrarily large, the distance above which they are separable converges to a finite value $D_{crit} = \mbox{lim}_{L\rightarrow \infty} D(L) = \mbox{lim}_{L\rightarrow \infty} C(L)L \sim 1$. The numerical results are stronger than our general proof from the preceding section as they imply the existence of a critical distance above which collective modes are separable for any choice of the detection profiles, while the analytical approach assures only that for any pair of profiles there exists a critical distance for finding entanglement between the corresponding collective modes. It is remarkable that our calculation indicates that this distance is of the order of the Compton wavelength.

\section{Results for the massive Klein-Gordon field}\label{sec:field}
\subparagraph{Asymmetric triangular profiles.}
We will follow the intuition on the optimal profiles obtained from the linear harmonic chain. The most profound feature of the optimal functions found in that case is their asymmetry (see Fig.\ \ref{profiles}) and the fact that they are mirror images of one another: $g_A(x) = g_B(-x)$. Thus, as a first approximation, to define the subsystems we use functions in the shape of asymmetric triangles. We restrict our numerical analysis to the one-dimensional case, however it is obvious that it may be extended to more dimensions. We parametrize each of the triangular profiles with the length of their support $L_{A/B}$ and the position of the tip $s_{A/B}\in (0,1)$, where $s_{A/B}=\frac{1}{2}$ gives a symmetric triangle. The property of mirroring results in both profiles having the same support size and tip position related by $s_B=1-s_A$, so that normalized profiles are given by
\begin{equation}\label{triangprof}
\begin{array}{ll}
g_{A}(x)\equiv g(s,L,x):=\left\{
\begin{array}{ll}
\sqrt{\frac{3}{L}}\left(\frac{x+L}{sL}\right) & \;\mbox{for}\;x\in (-L, -L(1-s)];\\
\sqrt{\frac{3}{L}}\left(\frac{-x}{(1-s)L}\right) & \;\mbox{for}\;x\in (-L(1-s), 0);\\
0 & \mbox{otherwise}. 
\end{array}\right. &
g_{B}(x)\equiv g(s,L,-x)
\end{array}
\end{equation}
These functions not only satisfy condition \eqref{decay} but also assumption A3. The latter is evident once we realize that Fourier transforms of the profiles are related by complex conjugation, $g_A(k)=\overline{g_B}(k)$, which is a direct consequence of the fact that the triangles are mirror images of one another. Notice that the profiles are chosen in such a way, that parameter $D=x_B - x_A$ is equal to the distance between their supports,  $D=0$ meaning neighboring triangles.
\begin{figure}[!ht]
\begin{center}
\includegraphics[width=5.5cm]{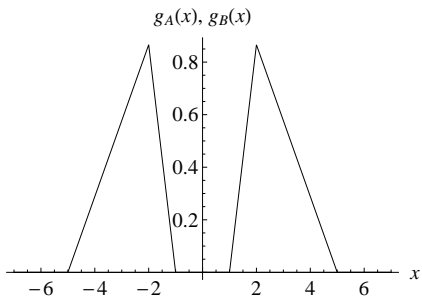}
\caption{Two mirroring triangular profiles for the support's size $L=4$, tip position $s=0.75$ and distance between the supports $D=2$.}
\label{triangles}
\end{center}
\end{figure}

Moreover, again, we set field's mass $m$ to 1 so that quantities of the length dimension are measured in Compton wavelengths. This finally makes the degree of entanglement \eqref{epsilon} depend only on three parameters: tip position $s$, size of the profiles' supports $L$ and their distance $D$. An exemplary setting is shown in the Fig.\ \ref{triangles}, which in our scheme corresponds to the situation where two bosonic modes defined by the profiles are associated with the regions of length 4 separated by interval of length 2 in Compton wavelength units.

\begin{figure}[!ht]
\begin{center}
\includegraphics[width=5.5cm]{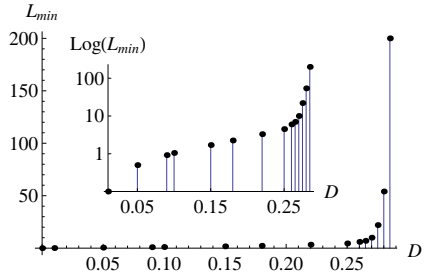}
\caption{Minimal length of the profiles' support $L_{min}$, for which entanglement appears, plotted as a function of their separation $D$. Inset: logarithmic plot of the same dependence.}
\label{pLmin}
\end{center}
\end{figure}

With this choice of profiles, entanglement is found numerically for a certain range of parameters. Here we present a summary of our results. First of all, for separation $D$ larger than $D_{crit}\approx 0.3$ the modes become separable -- no entanglement can be found for any choice of the remaining two parameters. From our analysis of the linear harmonic chain, the critical distance was estimated to be of order $1$. It was, however, done for \textit{optimal} profiles to which triangles are just an \textit{approximation}. For each separation $D < D_{crit}$ entanglement appears, once the size of the supports $L$ exceeds some minimal value $L_{min}(D)$. This minimal length increases with $D$. The  existence of the critical distance is manifested by $L_{min}(D) \rightarrow \infty $ for $D \rightarrow  D_{crit}$; see Fig.\ \ref{pLmin}.

\begin{figure}[!ht]
\centering
\begin{minipage}[r]{0.4\linewidth}
\includegraphics[width=5.5cm]{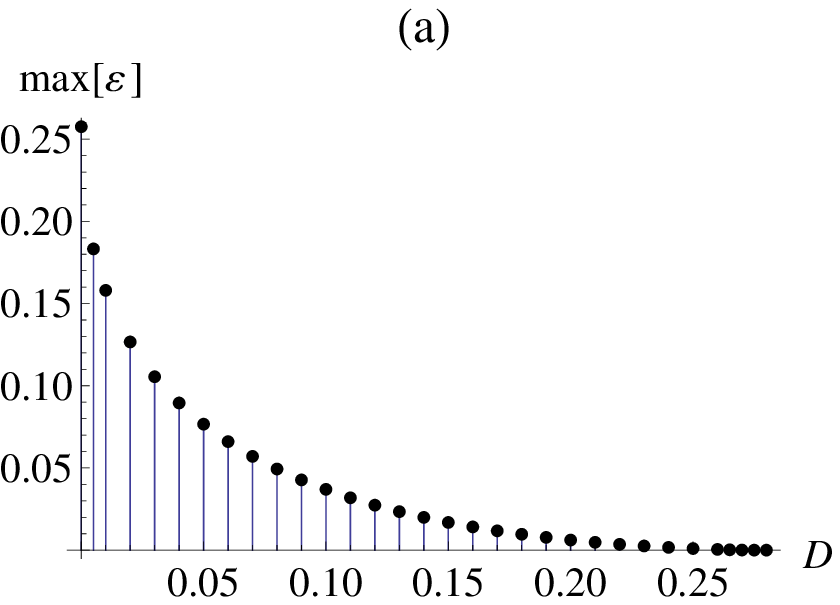}
\end{minipage}%
\begin{minipage}[l]{0.4\linewidth}
\includegraphics[width=5.5cm]{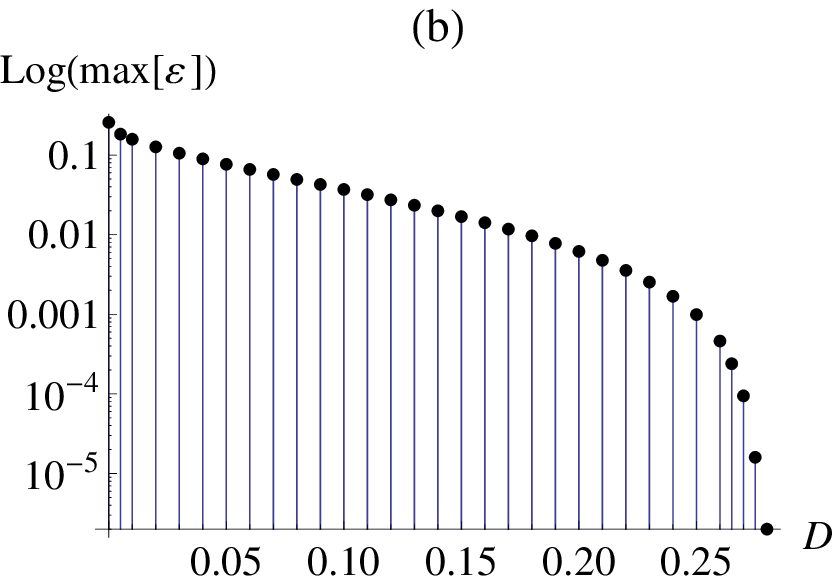}
\end{minipage}
\caption{(a) Degree of entanglement $\varepsilon$ maximized over the size of the profiles' support $L$ and the tip position $s$ as a function of the separation $D$, i.e. maximal available entanglement max[$\varepsilon$] as a function of the separation of the profiles $D$. (b): Logarithmic plot of the same dependence.}
\label{emaxd}
\end{figure}

For given separation $D$, we can maximize the entanglement measure \eqref{epsilon} over the size of the supports $L$ and tip position $s$. In this way we obtain the maximal available entanglement as a function of the subsystems' separation $D$ (Fig.\ \ref{emaxd}). For $D \rightarrow  D_{crit}$ entanglement goes to zero. From the logarithmic plot we infer that in the intermediate range of separation parameter values, the entanglement measure $\varepsilon$ optimized over the two remaining parameters ($L$ and $s$) decreases exponentially with the distance $D$.

\begin{figure}[!ht]
\centering
\begin{minipage}[r]{0.4\linewidth}
\includegraphics[width=5.5cm]{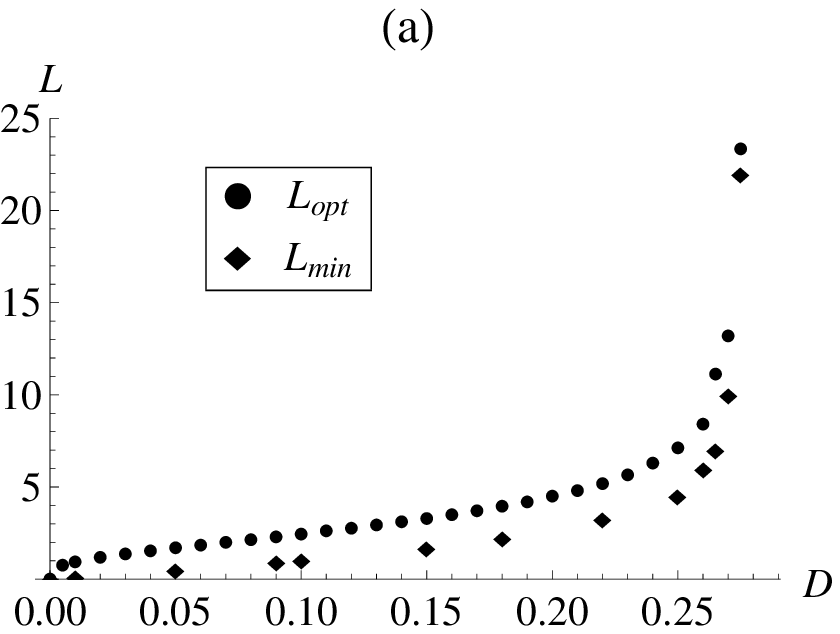}
\end{minipage}%
\begin{minipage}[l]{0.4\linewidth}
\includegraphics[width=5.5cm]{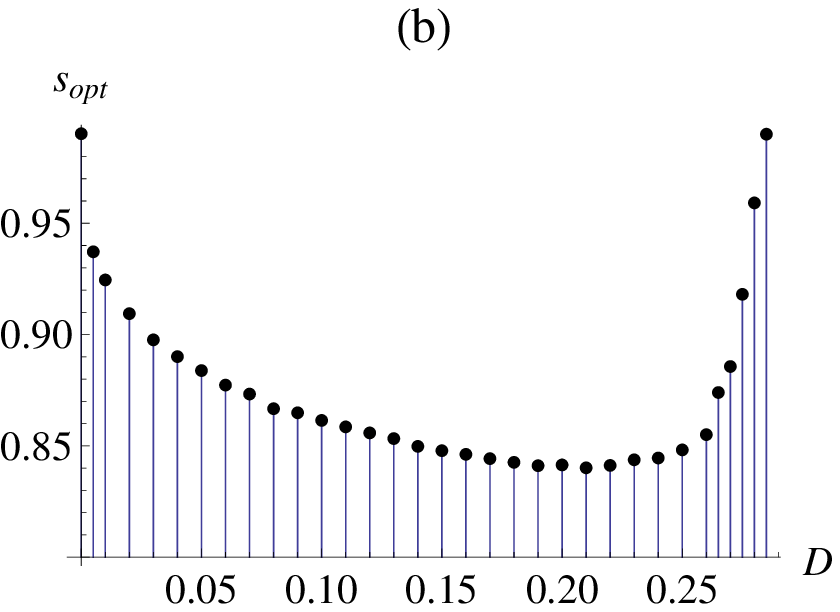}
\end{minipage}
\caption{Optimal and minimal length of the profiles' supports, plot (a), and optimal position of the triangles' tip, plot (b), as functions of the separation $D$ of the profiles' support. By optimal values of parameters we understand such that maximize our entanglement measure $\varepsilon $.}
\label{slopt}
\end{figure}

%
%
In Fig.\ \ref{slopt} we show the dependence of the optimal (maximizing entanglement) values of the size of the supports $L_{opt}$ and tip position $s_{opt}$. Qualitatively, the behavior of $L_{opt}$ and $L_{min}$ is the same. $L_{opt}$ is of the order of the Compton wavelength in the intermediate region of the separation parameter values. More interesting is the dependence of the optimal tip position $s_{opt}$ on the separation $D$. It reaches its minimal value $s_{opt}\approx 0.84$ for $D \approx 0.2$. Both for $D\rightarrow 0$ and $D\rightarrow D_{crit}$ the optimal tip position goes to 1. The difference between $L_{opt}$ and $L_{min}$ just confirms the result of \cite{kofler}, that entanglement may emerge by going to larger blocks.

\begin{figure}[!ht]
\centering
\begin{minipage}[r]{0.4\linewidth}
\includegraphics[width=5.5cm]{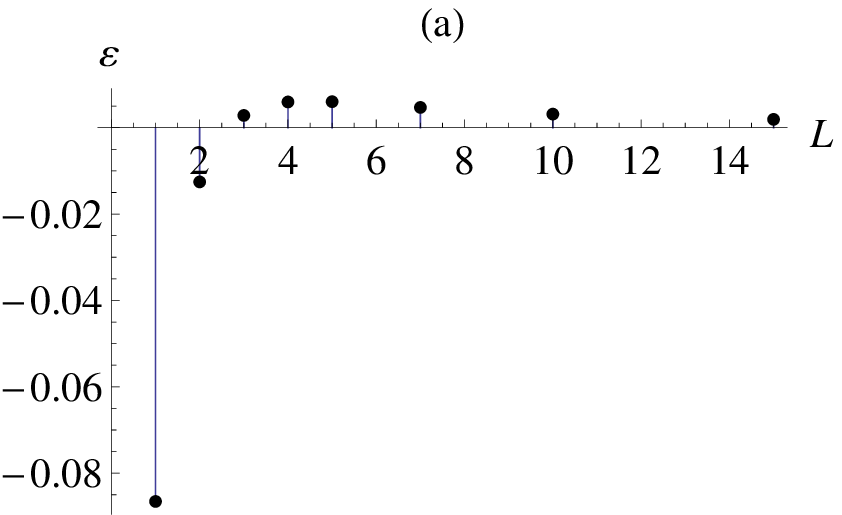}
\end{minipage}
\begin{minipage}[l]{0.4\linewidth}
\includegraphics[width=5.5cm]{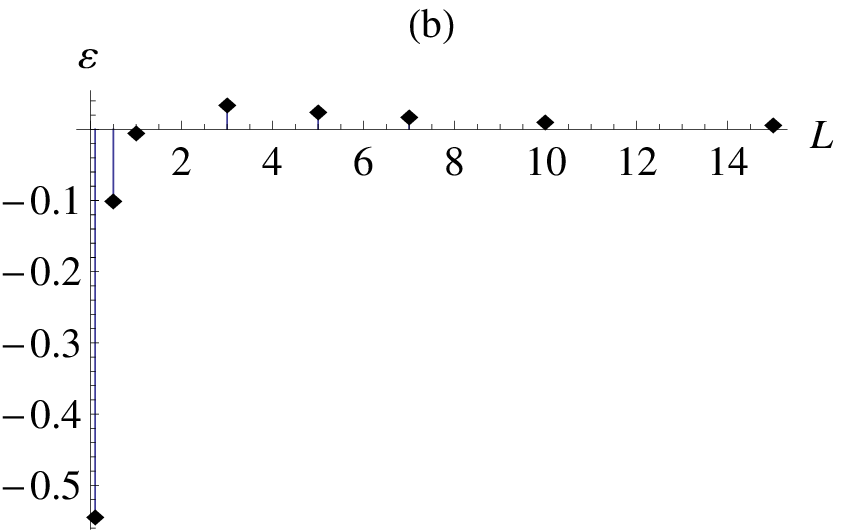}
\end{minipage}
\caption{Entanglement as a function of the size of the profiles' support $L$ for given separation of the profiles, $D=0.2$ for plot (a) and $D=0.1$ for plot (b). To obtain each point on the plots, the degree of entanglement $\varepsilon $ was maximized over the tip position $s$.}
\label{emaxld12}
\end{figure}

For the sake of completeness we include the plots of entanglement as a function of the size of the profiles' supports for separation $D=0.2$ [Fig.\ \ref{emaxld12} (a)] and $D=0.1$ [Fig.\ \ref{emaxld12} (b)]. To obtain each point of the plots, we optimized over the tip position.

\begin{figure}[!ht]
\centering
\begin{minipage}[r]{0.4\linewidth}
\includegraphics[width=5.5cm]{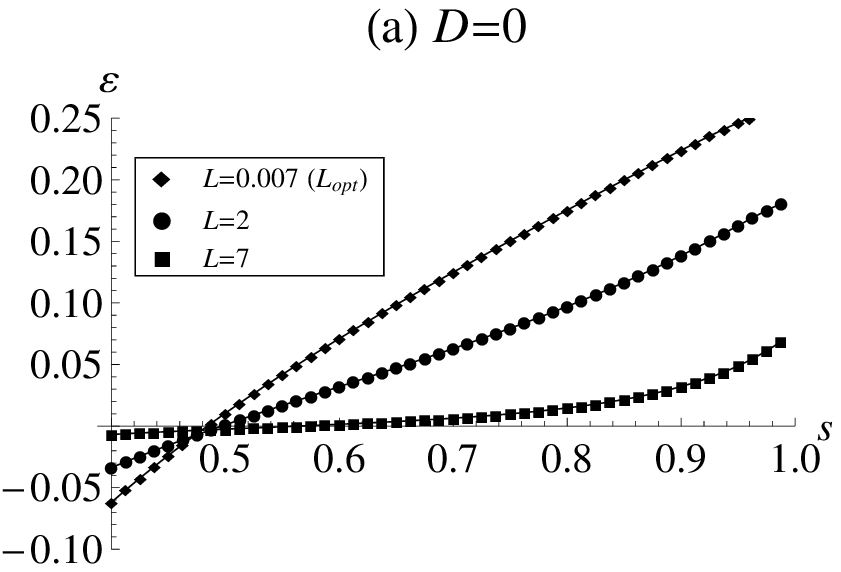}
\end{minipage}%
\begin{minipage}[l]{0.4\linewidth}
\includegraphics[width=5.5cm]{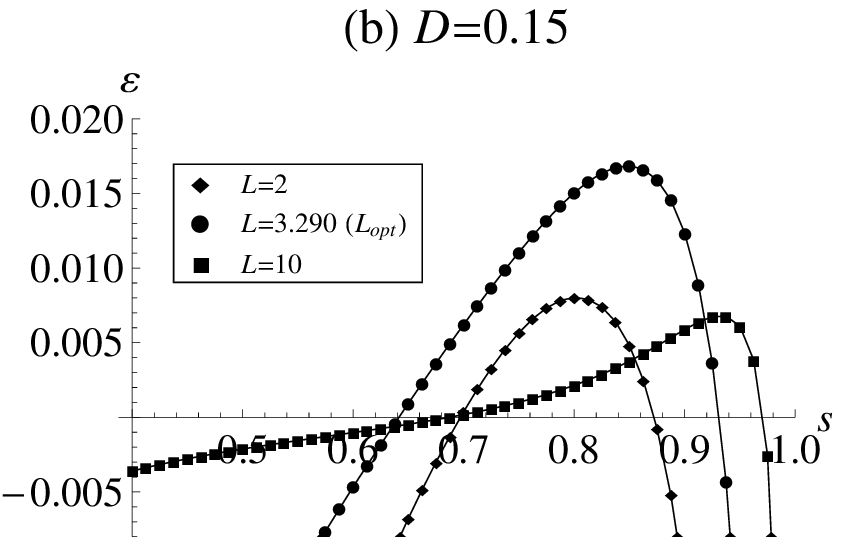}
\end{minipage}
\caption{Entanglement as a function of the tip position $s$ for a few chosen sizes of profiles' supports, $L$, and the separation of the profiles: $D=0$ for plot (a) and $D=0.15$ for plot (b). $L_{opt}$ indicates the value of $L$ parameter giving maximal entanglement for the relevant value of the separation parameter $D$, (compare Fig.\ \ref{slopt}).}
\label{emaxsld12}
\end{figure}

The larger the values of the separation are and the farther the size of the profiles' supports is from its optimal value (for given separation), the more sensitive our entanglement measure is to changes of the tip position. In other words: as the separation $D$ increases, $\varepsilon$ becomes more and more peaked over optimal values of $L$ and $s$ parameters. We exemplify this in Fig.\ \ref{emaxsld12}, where entanglement is plotted for separation $D=0$ [Fig.\ \ref{emaxsld12} (a)] and  $D=0.15$ [Fig.\ \ref{emaxsld12} (b)] as a function of the tip position for a few values of the size of the triangles' supports.

\begin{figure}[!ht]
\centering
\begin{minipage}[r]{0.4\linewidth}
\includegraphics[width=5.5cm]{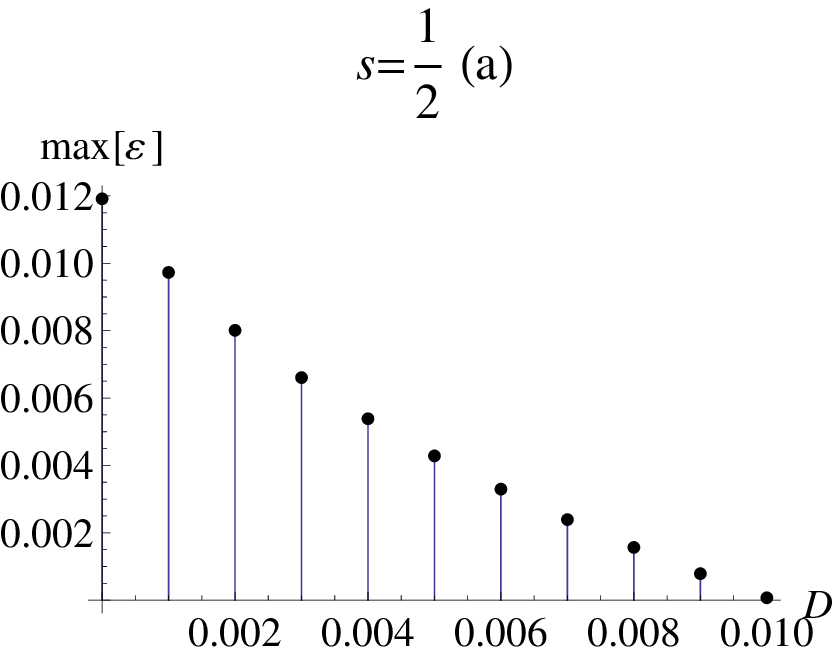}
\end{minipage}%
\begin{minipage}[l]{0.4\linewidth}
\includegraphics[width=5.5cm]{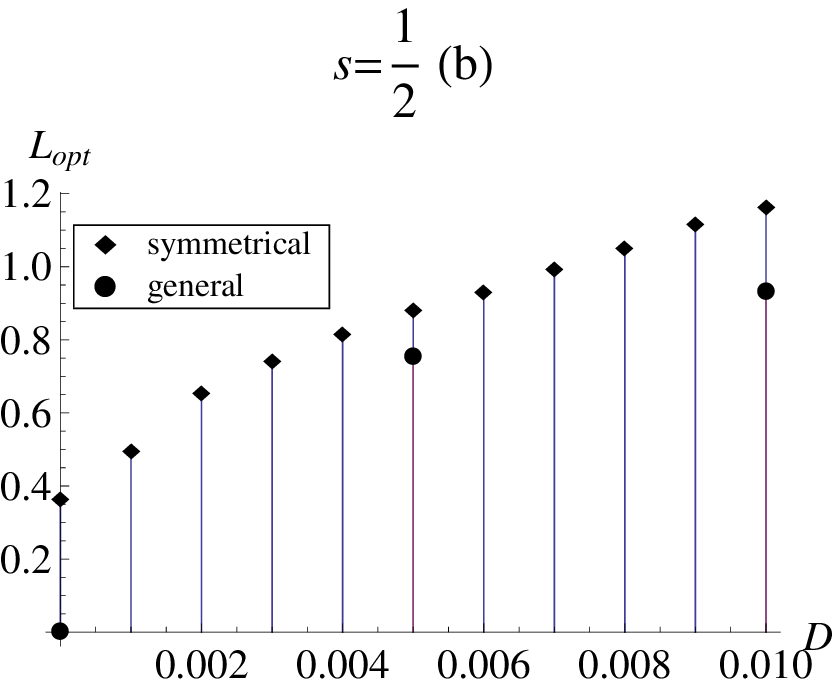}
\end{minipage}
\caption{Results for the symmetrical triangles, i.e.\ profiles \eqref{triangprof} with $s=\frac{1}{2}$. Plot (a): for given separation $D$, we maximize $\varepsilon$ over the size of the supports $L$ keeping $s=\frac{1}{2}$. In this way, we obtain the maximal available entanglement for symmetrical profiles as a function of the separation. Plot (b): optimal lengths of the symmetric and general profiles. By optimal we understand those that maximize the degree of entanglement $\varepsilon$.}
\label{symmplots}
\end{figure}

At the beginning of this section, we stressed that the important feature of the profiles defining entangled modes is their asymmetry. Naturally, there arises the question to which extent this property is crucial. Is asymmetry necessary to obtain entanglement? In general the answer is negative, however, as shown in Fig.\ \ref{symmplots} (a), the maximal amount of entanglement available in the case of symmetric profiles is more than 1 order of magnitude smaller than in general situation (i.e.\ when we can vary position of the tip of the triangles). Moreover, symmetrical modes become separable when the distance between their supports exceeds a value about 0.01 which is also more than an order of magnitude smaller than the critical distance estimated in the general, asymmetric, case ($D_{crit}\approx 0.3$). In Fig.\ \ref{symmplots} (b) we plotted the dependence of the optimal size of symmetric profiles as a function their separation. There are also included points from the Fig.\ \ref{slopt} (a), i.e.\ optimal lengths of the more general profiles for given separation. The optimal lengths of the symmetric profiles and of the general ones are, unlike the degree of entanglement, comparable. This result shows that the asymmetry of the profiles is indeed important for maximizing the degree of entanglement for given distance between the subsystems.  

We also stress that the exact shape of the triangle is not necessary. Since the space of compactly supported, smooth functions is dense in $L^p(\mathbb{R}^3)$ for $p\geq 1$, all the integrals in the expression for $\varepsilon$ obtained with the triangles may be approximated to arbitrary precision with smooth functions satisfying all demanded constraints. Also an explicit ``smoothing'' is possible with the use of a series of Gaussian functions  $f_a(x)=\frac{1}{a \sqrt{\pi}}e^{-\frac{x^2}{a^2}}$. They approach the Dirac delta in the weak limit i.e.\ $\lim_{a\to 0^+} \int_{-\infty}^{\infty}f_a(x)g(x) \, dx = g(0)$. Convolving triangles with the function $f_{\tilde a}(x)$ for $\tilde a$ sufficiently close to $0$, we obtain a smooth profile that, again, approximates the entanglement measure $\varepsilon$ calculated with triangular profiles up to arbitrary precision. The price is that canonical commutation relations are also satisfied approximately -- as Gaussians are not compactly supported, so are the final smoothened ``triangles.''

Finally, we would like to point out that the joint state of the two regions, which we here consider, is in principle mixed, so it may no longer be useful e.g. for quantum communication. As proved in \cite{distillability}, for Gaussian two-mode systems entanglement (i.e. $\varepsilon>0$) is equivalent to distillability of the state. However, it is impossible to distill entanglement utilizing two identical, symmetrical copies of a two-mode Gaussian state at a time and performing Gaussianity preserving operations supported by classical communication \cite{eisert} (non-Gaussian operations would be required). Entanglement distillation (or purification) is a transformation that consists of local operations and classical communication bringing several copies of a mixed entangled state into (approximately) pure entangled states  which can further be utilized for quantum communication and quantum computation. (See Sec.\ \ref{sec:collective} for a justification that the state we define with the detection profiles is indeed Gaussian.) 

\section{Conclusion}\label{sec:conclusion}
In this paper, we investigated entanglement between two regions of a scalar Klein-Gordon field in the vacuum state. By spatially integrating over field operators (and conjugate momenta) with two real, compactly supported functions we defined two pairs of collective phase space operators representing two bosonic modes, i.e.\  - subsystems corresponding to the regions given by the supports of the functions.
Reducing the vacuum of a scalar field to these particular modes, we studied entanglement between them with the use of an entanglement measure for two-mode continuous variables states (based on \cite{simon}).

For every two functions satisfying the aforementioned constraints (see Sec.\ \ref{sec:collective}), it is proved that the corresponding subsystems are separable if the distance between them (i.e.\ between the supports of the functions) is larger than some finite value. From a numerical analysis of the discretized Klein-Gordon field we obtained strong  indication that all considered modes become separable for separations larger than 1 Compton wavelength. We also gave an explicit example of a pair of functions (asymmetric triangles) that define entangled modes and investigated numerically the amount of entanglement in the corresponding system.

The approach presented here aimed to take into account limitations that real experiments put on the properties that can be measured. From this point of view, our definition of observables is a reasonable first approximation towards a fully realistic treatment of the problem of vacuum entanglement as it assumes that only localized collective operators can be observed (since we cannot resolve field operators in single space-time points). Also an interaction has been proposed that could implement the desired measurements.

According to our numerical analysis of the discretized field, in half spaces separated by more than a Compton wavelength we cannot find any entangled modes defined with the considered functions. In \cite{werner}, a violation of Bell's inequalities in the vacuum is in principle possible at arbitrary separations. Although our result cannot be directly compared with works on Bell's inequality violations (because our observables are Gaussian and on a Gaussian state such operators cannot show violation of Bell's inequalities \cite{speakable}), we would like to point that the algebra of observables considered in \cite{werner} is much richer than ours. This implies that the operators needed for this violation (in the large separation regime) must be of a more intricate form than proposed here ( e.g.\ involving higher powers of the field operators). On the other hand, as the authors of \cite{werner} comment, if the distance $D$ between the regions probed is much larger than a few Compton wavelengths of the lightest particle in the theory then the maximal Bell violation in the vacuum will necessarily be too small to be observed\footnote{In their approach, the parameter describing the maximal violation of Bell's inequalities with operators from two local algebras, assigned to space-time regions separated by $D$, decays exponentially with $m\,D$ (where $m$ stands for the mass of the lightest particle in the theory).}. This shows that even with a wide range of allowed observables vacuum entanglement, if at all accessible, should be tested at small distances (of the order of the field's Compton wavelength\footnote{The fact that entanglement is only a short distance property of the vacuum is further confirmed by the fact that it is possible to transform the vacuum into a separable state by means of a nonlocal unitary involving only points at a Compton wavelength distance \cite{papero}.}). Our result asserts that, once we can probe close enough regions, to access vacuum entanglement we can  restrict our measurements only to very simple observables, namely, field (and conjugate momentum) operators averaged over spatial regions. Such observables are usually considered as simplest to implement, at least in optics. 

There are several possible generalizations of the presented approach. First of all, each phase space operator ($\hat Q_{A/B}, \hat P_{A/B}$) may be defined via a different profile and the restrictions given by assumptions A1-A3 may be abandoned. Both these situations are not covered by our considerations. Further, we may allow for more general observables than field operators averaged over spatial regions, as e.g.\ in \cite{werner} or by considering space-time regions. Also investigating entanglement between excitations of local Hamiltonians (i.e.\ restricted to some space region) in the global vacuum state may be an interesting line of research.

Our results have importance for investigating whether the vacuum of a quantum field has any operational meaning and if it may be accessed as an entanglement resource. Moreover, the method here presented is directly applicable to other than vacuum bounded energy states and also to systems described within nonrelativistic quantum field theory. The most extreme example of the latter (if the range of the correlations is considered), the BEC state, will be studied elsewhere. 

\begin{acknowledgments}
We thank Janet Anders, Piotr Kosi\'{n}ski, Federico Piazza and Reinhard Werner for insightful discussions and helpful remarks. M.Z would like to acknowledge the Erwin Schr\"{o}dinger International Institute for Mathematical Physics, Vienna, Austria for the JRF Scholarship during 1.III-30.VI 2009. This work was supported by the Austrian Science Foundation FWF (SFB FoQuS, Project No. P19570-N16, and CoQuS) and the European Commission through Project QAP (No. 015846).
\end{acknowledgments}

\end{document}